# Additive manufacturing of solid diffractive optical elements via near index matching

Reut Kedem Orange[1,2], Nadav Opatovski[1,2], Dafei Xiao[1,2], Boris Ferdman[1,2], Onit Alalouf[2,3], Sushanta Kumar Pal[4], Ziyun Wang[5,6], Henrik von der Emde[7], Michael Weber[7], Steffen J. Sahl[7], Aleks Ponjavic[5,6], Ady Arie[4], Stefan W. Hell[7,8], Yoav Shechtman[1,2,3,†]

[1]Russell Berrie Nanotechnology Institute, Technion—Israel Institute of Technology, Haifa, Israel
[2]Lorry Lokey Interdisciplinary Center for Life Sciences and Engineering, Technion—Israel Institute of Technology, Haifa, Israel
[3]Department of Biomedical Engineering, Technion—Israel Institute of Technology, Haifa, Israel
[4]School of Electrical Engineering Fleischman Faculty of Engineering, Tel Aviv University, Tel Aviv 69978, Israel
[5]School of Physics and Astronomy, University of Leeds, Leeds, UK
[6]School of Food Science and Nutrition, University of Leeds, Leeds, UK
[7]Department of NanoBiophotonics, Max Planck Institute for Multidisciplinary Sciences, Göttingen, Germany
[8]Department of Optical Nanoscopy, Max Planck Institute for Medical Research, Heidelberg, Germany

[†] Corresponding author: yoavsh@technion.ac.il

## Abstract

Diffractive optical elements (DOEs) have a wide range of applications in optics and photonics, thanks to their capability to perform complex wavefront shaping in a compact form. However, widespread applicability of DOEs is still limited, because existing fabrication methods are cumbersome and expensive.

Here, we present a simple and cost-effective fabrication approach for solid, high-performance DOEs. The method is based on conjugating two nearly refractive index-matched solidifiable transparent materials. The index matching allows for extreme scaling up of the elements in the axial dimension, which enables simple fabrication of a template using commercially available 3D printing at tens-of-micrometer resolution. We demonstrated the approach by fabricating and using DOEs serving as microlens arrays, vortex plates, including for highly sensitive applications such as vector beam generation and super-resolution microscopy using MINSTED, and phase-masks for three-dimensional single-molecule localization microscopy.

Beyond the advantage of making DOEs widely accessible by drastically simplifying their production, the method also overcomes difficulties faced by existing methods in fabricating highly complex elements, such as high-order vortex plates, and spectrum-encoding phase masks for microscopy.

# Introduction

Diffractive optical elements (DOEs) enable highly complex shaping of light by a single, compact, optical component. This capability makes DOEs attractive in a variety of applications[1], including aberration correction[2], augmented reality[3], imaging systems[4,5], solar energy[6], sensitive microscopy[7–9] and more[10,11]. Despite their advantages, fabrication of DOEs can be extremely challenging as high-quality light-shaping entails fabrication with sub-wavelength precision. Complicated DOEs often consist of topology at the tens-of-nanometer resolution, which typically necessitates fabrication methods that are expensive, time consuming, and require special infrastructure[1,12], e.g. photolithography or direct machining methods. Moreover, fabrication of highly-complex elements by binary photolithography is fundamentally limited due to per-step accumulated error and practically-bounded number of possible etching steps; grayscale-photolithography, which overcomes some of these limitations, is typically complicated by high sensitivity to material and process parameters and the requirement for complex illumination or scanning[1,13].

Additive manufacturing (AM) is a versatile and efficient fabrication method that offers quick fabrication times and cost-effective production for intricate components[14]. In addition, the AM process is relatively robust to the geometric complexity of the manufactured parts. Although AM has recently showed promising results in printing optical components[15–18], outstanding limitations that restrict widespread usability of AM for optical fabrication still exist. These limitations relate to surface quality, uniformity of the optical properties e.g., refractive index, transparency and a fundamental tradeoff between precision, component size and fabrication duration. Recent work has demonstrated that the concept of index-matching can be used to overcome the stringent fabrication constraints associated with traditional DOE fabrication, while maintaining high performance and functionality[19]. In this demonstration, a 3D printing-based polymer template was placed inside a nearly index-matched liquid bath; However, immersing the DOE in a liquid limits its wide applicability as an element to be integrated into any optical system.

Here we present a fast and simple method to fabricate fully solid DOEs utilizing all the advantages AM offers. Our method uses commercially available AM to fabricate high-quality DOE reducing cost and fabrication time by orders of magnitude. This is done by conjugating two near-index matched materials, effectively scaling up the critical dimension of the DOE from the nano to the micro scale and correspondingly increasing tolerance for fabrication error. In this concept, the requirement for fabrication precision is traded for precision in the refractive index of the materials, which is relatively easy to obtain at the relevant scales. We demonstrate the usefulness and versatility of our method by fabricating and using DOEs from a variety of optical fields, including: a Fresnel microlens array, phase masks for 3D super resolution microscopy, encoding axial[7] and spectral[20] information of fluorescence sources, and spiral phase plates (SPP) generating vector beams with different topological charges[21], including their application in MINSTED nanoscopy[8].

## Results

A DOE can be regarded as 2.5D pixelated surface, where the height of each pixel corresponds to the thickness of the transparent material. The wavefront shaping effect of the DOE can be described by the accumulated phase of light propagating through a pixel, given by:

$$(1) \Delta \phi_{(i,j)} = \frac{2\pi}{\lambda} h_{(i,j)} \Delta n ,$$

where $\Delta \phi_{(i,j)}$ is the accumulated phase in pixel (i,j), $\lambda$ is the wavelength, $h_{(i,j)}$ is the height of pixel (i,j), and $\Delta n = (n_{DOE} - n_{env})$, where $n_{DOE}, n_{env}$ are the refractive indices of the DOE material and the environment surrounding it, respectively. Typically, DOEs are made of high-quality glass or polymer[22,23] with refractive index of $n_{DOE} \approx 1.5$, and are used in an air-filled environment such that $n_{env} = 1$, therefore $\Delta n$ is around 0.5. Under these circumstances, typical DOEs span a thickness range of a few micrometers at most, and require precision on the order of tens of nm.

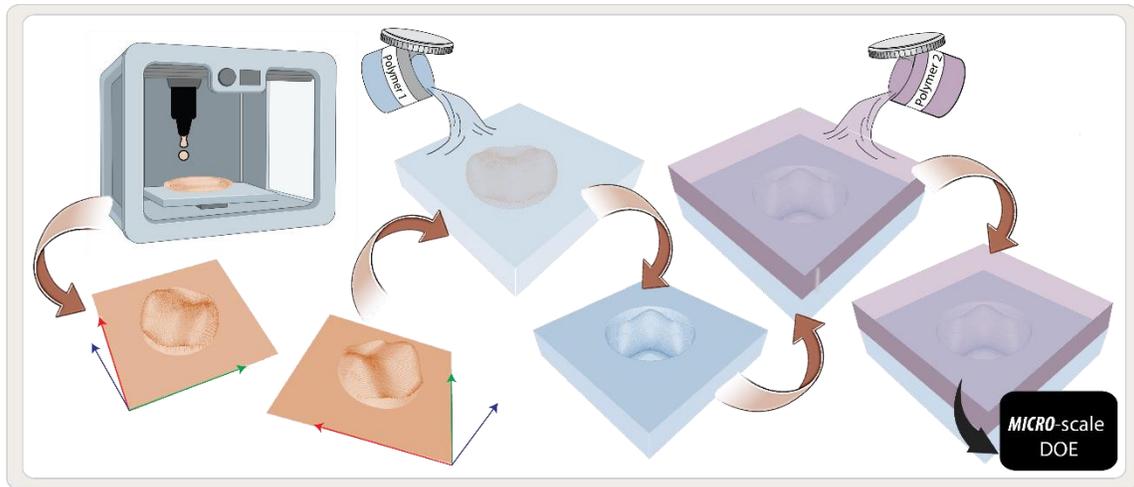

**Fig. 1: Fabrication method.** An illustration of the main steps of the fabrication process of the near index matched solids DOE. A 3D template is printed using commercially available 3D printing. The template is converted to a first transparent layer using a first polymer. Next, The first transparent layer is extracted from the template and a second polymer is polymerized above the first layer for achieving the final solid DOE.

Here, we scale up the axial DOE dimension by 2-3 orders of magnitude, by reducing the $\Delta n$ drastically from ~0.5 to $0.001 \leq \Delta n \leq 0.007$. This is done by conjugating two near index matched materials, containing the desired DOE profile in the interface between them, as described below.

The first step of our fabrication process is to print a micro-scale template using commercially available AM. Next, we convert the DOE profile to a first transparent layer by casting the first polymer on the template and polymerizing it. After extracting the template, a glass is attached to the flat side of the first layer using plasma treatment. The final step is to cast the second polymer above the transparent profile of the first layer, attach a second glass above it and polymerize it, to ensure the flatness of the DOE in the interface with air (Fig.1).

To demonstrate the versatility of the fabrication method, we produced and applied various DOEs.

First, we fabricated a Fresnel micro-lens array (MLA) with a high fill factor (100%)[24]. Micro lens arrays are used in a wide range of applications, from sensing to solar energy and microfluidics for focusing, imaging, and beam-shaping[25,26]. Here, we fabricated a 3X3 MLA made of Fresnel square lenses, each 2 mm X 2 mm in size which focuses an incident Gaussian beam to an array of spots. The obtainable spot size is 340 $\mu m$, which is close to diffraction limited for this geometry ($2\frac{\lambda f}{d}$ ~320 $\mu m$) (Fig. 2). Camera (Pixelink, PL-D7512MU-T) pixel size is 3.45 $\mu m$.

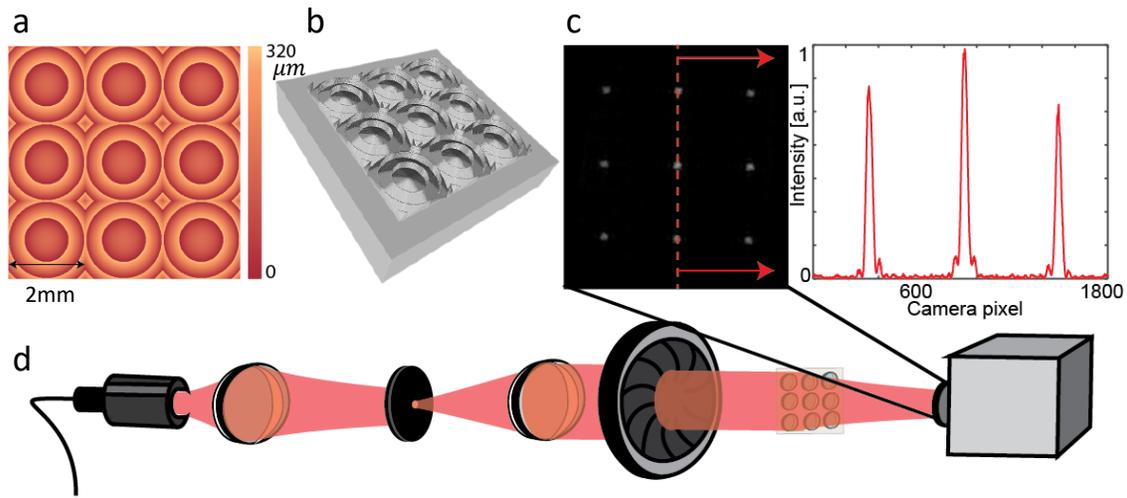

**Fig. 2 Micro lens array**. a) A 2D height map and b) 3D illustration of the MLA c) the acquired image on the camera showing the spots focused by the MLA as well as the intensity plot of three spots. d) The optical system used for MLA demonstration.

Next, we demonstrated the fabrication and application of spiral phase plates (SPPs). A SPP modifies the phase profile of an incident Gaussian beam by adding a phase shift that varies linearly with the azimuthal angle around the center of the plate, resulting in a spiral phase pattern. The SPPs can be used to create 'donut' shaped illumination, i.e. a ring of azimuthally uniform intensity with a dark spot in the center. The ring diameter and the corresponding size of the dark spot are determined by the topological charge of the SPP (*m*), which corresponds to the number of phase cycles in a complete turn around the element center. Controlling the topological charge using our method is obtained relatively simply, by adjusting the polymer parameters (changing Δ*n*) for a given wavelength.

Importantly, our method enabled us to easily produce high-order SPPs, which are difficult to obtain using existing fabrication methods; by exploiting phase wrapping in a 'pizza'- like pattern (Fig. 3 a) we demonstrated SPPs with topological charges of *m*=8 (Fig. 3 b) and *m*=16 (Appendix Fig. 7). The topological charge was measured using a cylindrical lens-based method[27] (Fig. 3b).

We demonstrated the generation and characterization of vector field singularities (V-points) using SPPs fabricated by our method. Vector beams are widely used for both fundamental and applied studies, for applications including trapping and manipulation of micro-particles, super-resolved imaging, laser materials processing, as well as classical and quantum optical communication[28]. This motivated us to realize such beams by using our fabricated SPPs.

V-points are the isolated, stationary points in a spatially varying linearly polarized field where the orientation of the electric field vector is undefined. They are characterized by a topological parameter, namely the Poincare–Hopf index $\eta = \frac{1}{2\pi}\oint \nabla\gamma \cdot dl$, where $\gamma$ is the azimuth of the polarization ellipse. Radial and azimuthal polarizations are examples of first-order V points with η=1. These singularities can be generated by the superposition of equal and opposite orbital angular momentum states in orthogonal circular polarization basis. The expression for V-point singularities in circular basis is:

$$(2)\vec{E}_V = r^{|m|} \cdot e^{\frac{-r^2}{w^2}} \cdot \{A_1 e^{im\varphi}\hat{e}_R + A_2 e^{(-im\varphi+i\theta_0)}\hat{e}_L\}$$

Here, $A_1$ and $A_2$ are the amplitude scaling factors, $m$ is the topological charge of the vortex beam. The parameters $\varphi$ and $\theta_0$ correspond to azimuthal phase and relative phase difference between the orthogonal components respectively. The right and left circular unit basis vectors are represented by $\hat{e}_R$ and $\hat{e}_L$ respectively.

We experimentally realized V-points by using a modified Mach-Zehnder type interferometer (Fig. 3 c). A spatially filtered collimated 45 degree linearly polarized beam was divided into equal intensities of x and y-polarized light by a polarizing beam splitter. This was approximated as plane beams that were converted into helical beams by using a SPP, combined by a beam splitter. A quarter waveplate at 45 degrees was used to transform these x and y-polarized helical beams into right and left circularly polarized vortex beams respectively. Various vector beams can be realized by appropriately selecting the topological charge (*m*) of the spiral phase plates. The quarter waveplate and polarizer before the lens were used to measure different intensities, from which the Stokes parameters and the desired polarization distributions were extracted (Appendix Fig. 8).

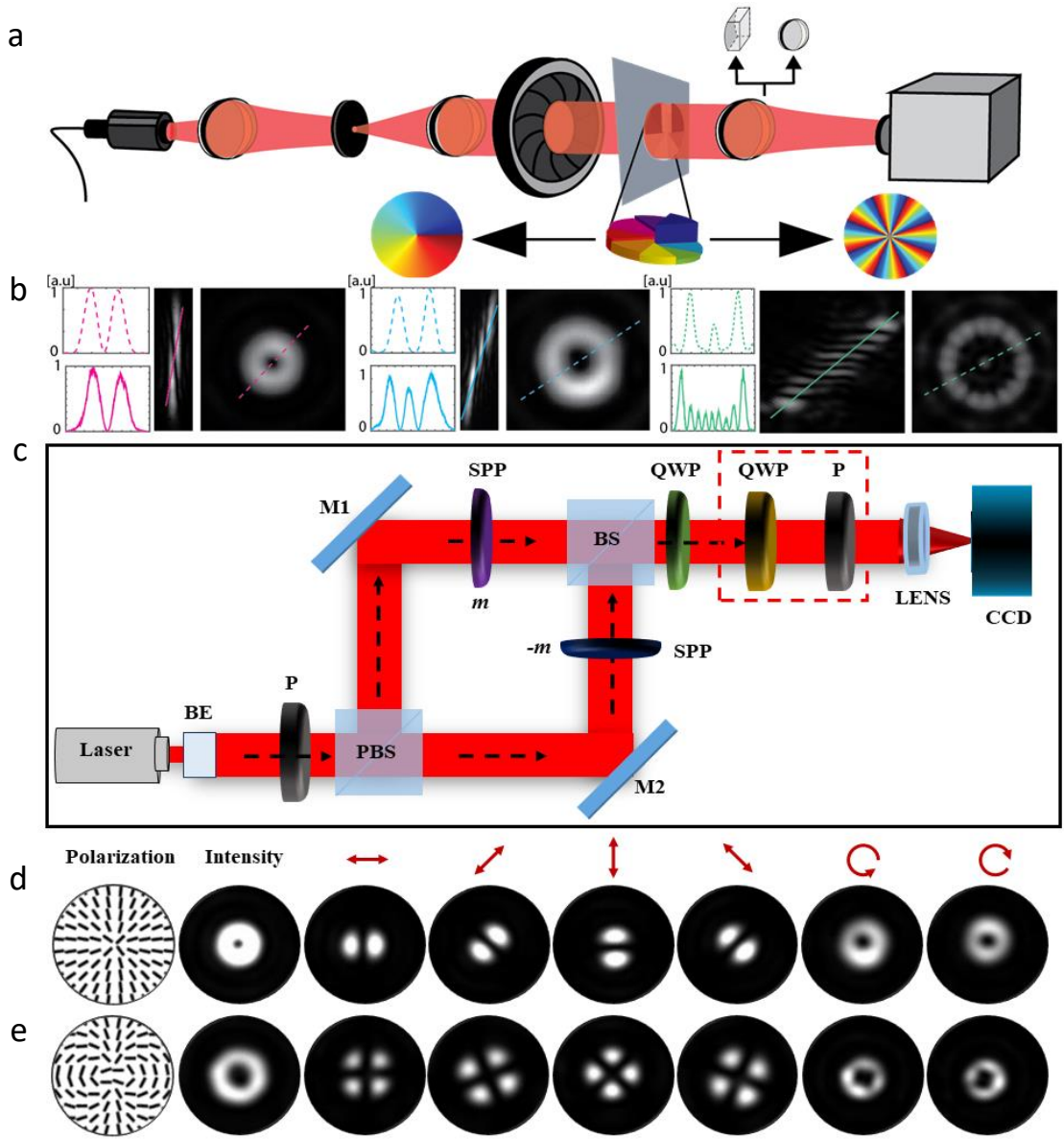

**Fig 3. Spiral phase plates.** a) Experimental setup for measuring the topological charge (*m*) and the profile intensity of the SPPs b) Experimental results of imaging a Gaussian beam passing through spherical or cylindrical lenses (dashed and continuous lines) and SPPs with different topological charges; m=1,2,8 from left to right respectively c) Experimental setup for generating vector field singularities. BE: Beam Expander; spatial filter assembly, P: Polarizer, M1, M2: mirrors, SPP: spiral phase plates, QWP: quarter-wave plate, PBS: polarizing beam splitter, BS: beam splitter, CCD: camera. d) and e) Experimental results showing the generation of vector field singularities; each row corresponds to a different set of experimentally measured intensity profiles generated with *m*=1 and *m*=2, respectively (right), and the theoretical distribution of the vector field (left), which is similar to the reconstructed distribution obtained from the experimental data (Appendix Fig. 8)

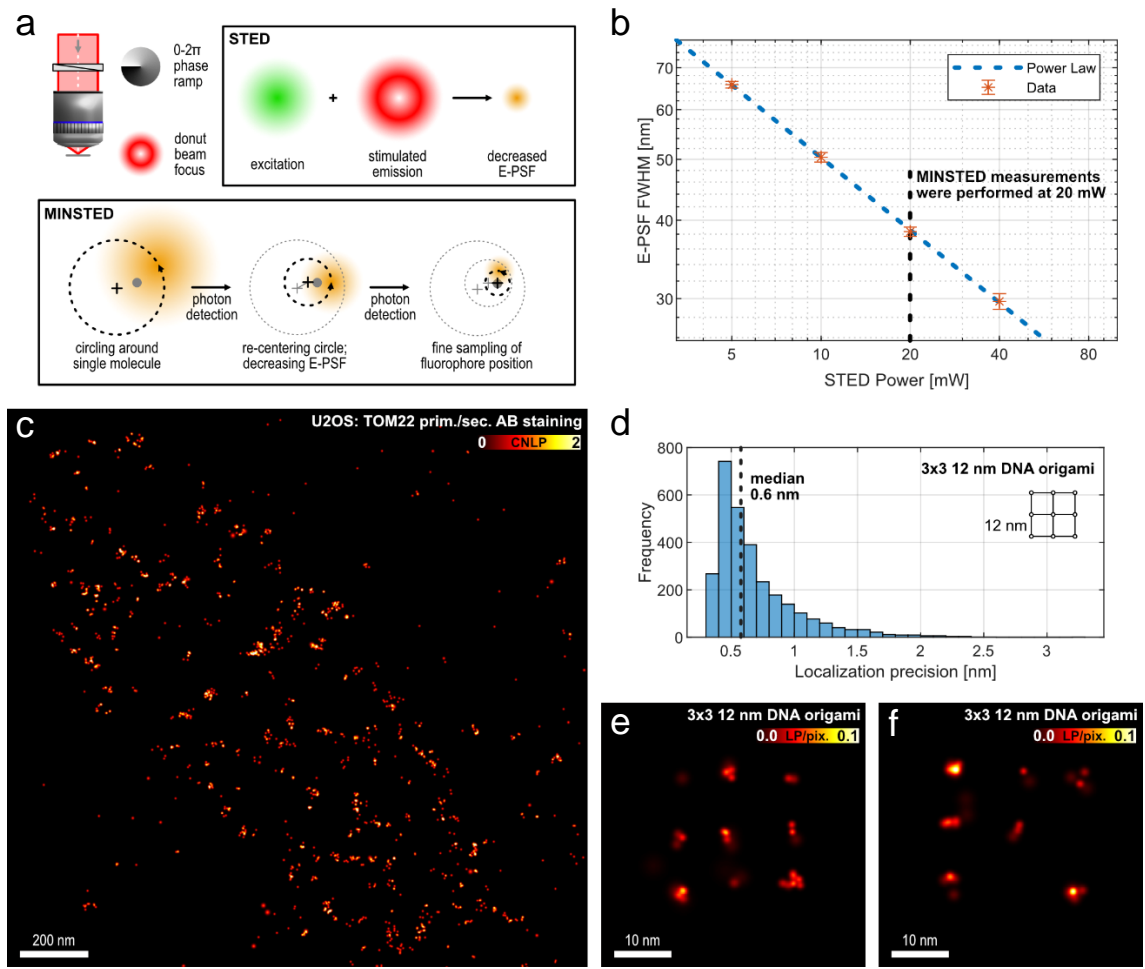

**Fig. 4 MINSTED fluorescence nanoscopy at ~1 nm localization precision.** In both STED and MINSTED nanoscopy, a mask encoding a 0-2π phase ramp can be used to produce a doughnut-shaped focal pattern featuring a "zero" (minimal) intensity point in its center for fluorescence inhibition. a) Principles of operation of STED and MINSTED imaging, featuring an effective point spread function (E-PSF) that can be tuned by the intensity in the STED inhibition pattern. b) Full width at half-maximum (FWHM) of the E-PSF as a function of STED power (636-nm wavelength). The E-PSF was repeatedly measured with immobilized single Cy3B fluorophores. A power-law fit to the data is shown (dashed line). c) MINSTED recording of localizations of TOM22, a protein in the outer mitochondrial membrane, in U-2 OS cells stained with primary and secondary antibodies. Rendering was performed by displaying each localization as a Gaussian with amplitude unity and a standard deviation matching its localization precision (including saturation for highly precise localizations). We denote this quantity as the cumulative normalized localization probability (CNLP) (d-f) Localizations of 3×3 DNA origami grids with 12-nm spacing between fluorophore binding sites. d) Histogram of localization precision from single-molecule trace analysis, indicating a median precision (st. dev.) of 0.6 nm. e,f) Two examples of DNA origamis. Rendering was performed by displaying each localization as a Gaussian with a pixel-sum of unity and a standard deviation matching its localization precision (including saturation for highly precise localizations). The image thus encodes the localization probability per pixel (LP/pix.) Scale bars: 200 nm (c), 10 nm (e,f)

Next, we demonstrated the applicability of 3D-printed elements in highly-demanding scenarios, requiring particularly high photon efficiency. We first applied our SPP inside a

STED[29,30] instrument, and demonstrate super-resolution microscopy at performance equivalent to commercial lithographically fabricated SPPs. Recent advances have propelled STED imaging to the single-digit nm scale, with the advent of a concept called MINSTED[8,31]. MINSTED operates on single fluorophores and achieves their efficient localization with a photon-detection-guided targeting that continually decreases the FWHM of the so-called effective point spread function (E-PSF) resulting from the STED process and re-centers it towards the most probable position of the targeted molecule[8].

The high-resolution performance of all STED methods crucially relies on the production of a focal pattern of the STED light featuring a "zero" (minimal) intensity point (Fig. 4 a). Incorporating the SPP for ~630 nm operation into a recently reported MINSTED nanoscopy instrument[31], we performed recordings of single fluorophores, model structures and protein distributions in mitochondria of human cells with an excitation wavelength of 560 nm and a STED wavelength of 636 nm, at a repetition rate of 40 MHz. The fluorescence switching to yield individual fluorophores for localization was implemented by means of DNA-PAINT[32] as done previously[31]. The obtained resolution characterization data on individual molecules (Fig. 4b) clearly support a residual intensity in the zero point of <1% of that at the donut crest, i.e. on par with the best phase masks utilized for this purpose. The mask operates within the MINSTED setup at leading-edge performance levels, enabling sub-1-nm median localization precision (Fig. 4d) for DNA origami model structures and in cellular nanoscopy of mitochondrial proteins (Fig. 4c and e-f).

Next, we fabricated several DOEs for point spread function (PSF) engineering[7,9] and demonstrated their applicability in single-molecule 3D localization microscopy. In PSF engineering, the emission path of a standard microscope is expanded using a 4f system, and a DOE, namely, a phase mask, is placed in the Fourier plane. This enables the encoding of useful information such as depth or color in the shape of the PSF[7,20]. The double helix (DH) PSF[9], for example, encodes the depth of a point source by generating a PSF containing two lobes that rotate as a function of the emitter's axial position (Fig. 5 a). We used a DH mask, fabricated using our method, to perform 3D DH super-resolution imaging[33] of cell membrane topography of Jurkat T cells using resPAINT, for quantification of microvilli (finger-like protrusions) prevalence as well as determination of their size distribution[34] (Fig. 5).

Finally, we show that beyond our method's advantage in drastically simplifying DOE production, it can produce DOEs with extremely complex architectures, including large height ranges and challenging aspect ratios. Such an example is a multicolor Tetrapod phase mask; this phase mask encodes both the depth (over a ~3.5 µm range) and the color (green and red) of fluorescence emitters, in the shapes and the orientations of their PSFs, respectively[20] (Fig. 6 b). This mask alleviates the need for spectral channel-splitting, since imaging is done on a single optical channel, and spectral separation is achieved by image analysis. Such simultaneous depth and color encoding has only been demonstrated in the past using a liquid crystal spatial light modulator (SLM)[20]; this is due to the difficulty in fabricating a DOE with such highly demanding geometry. The limited photon efficiencies of liquid crystal SLMs, however, precluded its use for 3D + multicolor single molecule microscopy, which we demonstrate here for the first time. Additionally, we added to the design a phase-grating that only affects green fluorescence, which translate to a linear shift in the image plane and creates a lateral shift between the green and red fluorescence emitters. This feature is useful when tracking two close fluorescence emitters labelled with two different colors[35], because the shift between the two colors minimizes PSF overlap and simplifies image analysis.

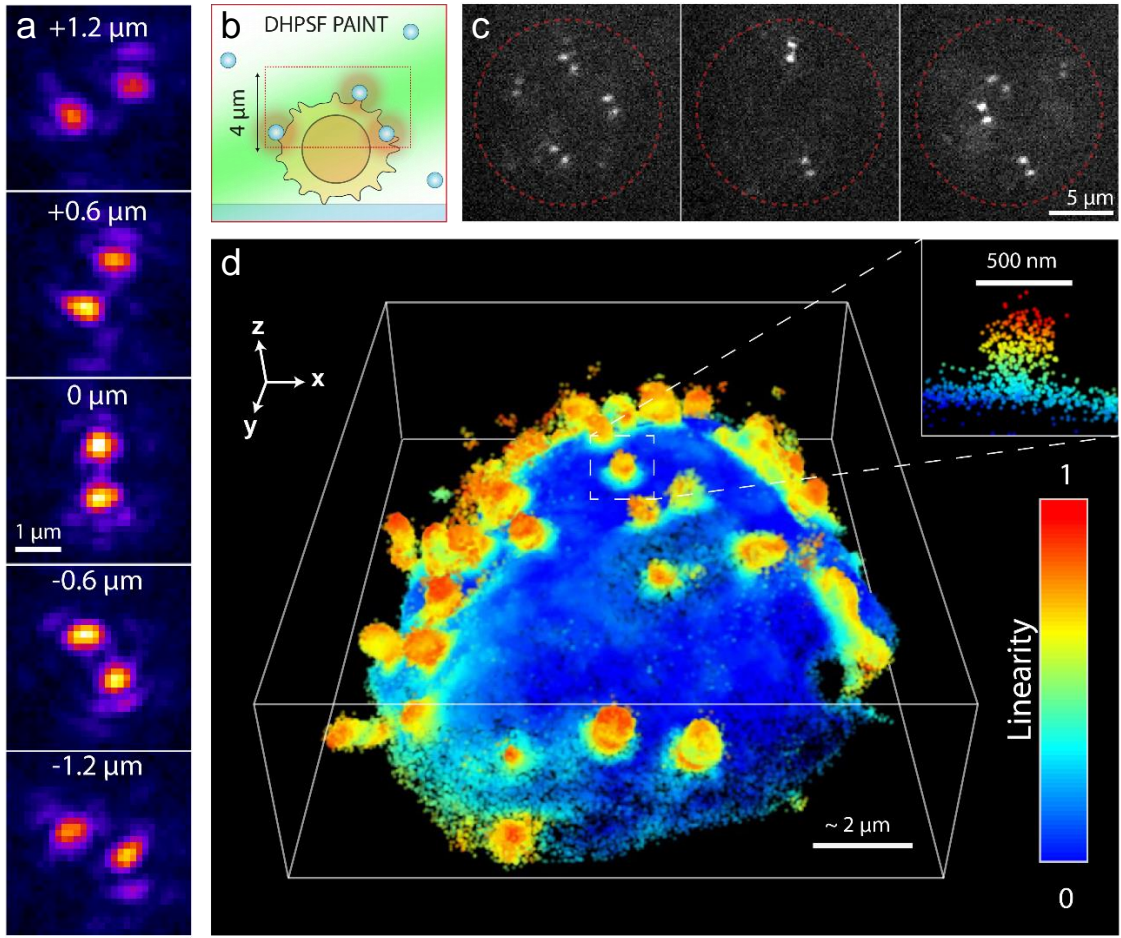

**Fig. 5: 3D super-resolution imaging of cell membrane topography using resPAINT with DH PSF.** a) DHPSF calibration for 100 nm fluorescent bead captured using 1.32 NA silicone oil objective. b) Cartoon showing large DOF imaging of the apical surface of a Jurkat T Cell with resPAINT. c) Representative frames of WGA-HMSiR binding to fixed Jurkat T cells (circle highlights cell) at 9.4 pH. (d) 3D super-resolution image of the apical surface of the cell membrane acquired using resPAINT and DHPSF. The image comprises ~300k localisations, collected over 300k frames with 30 ms exposure time. To highlight the topography of the cell membrane, we compared each data point to its local neighbours (200 nm radius shell) to quantify the curvature. This was done using principal component analysis, which is routinely used to classify point cloud data[42]

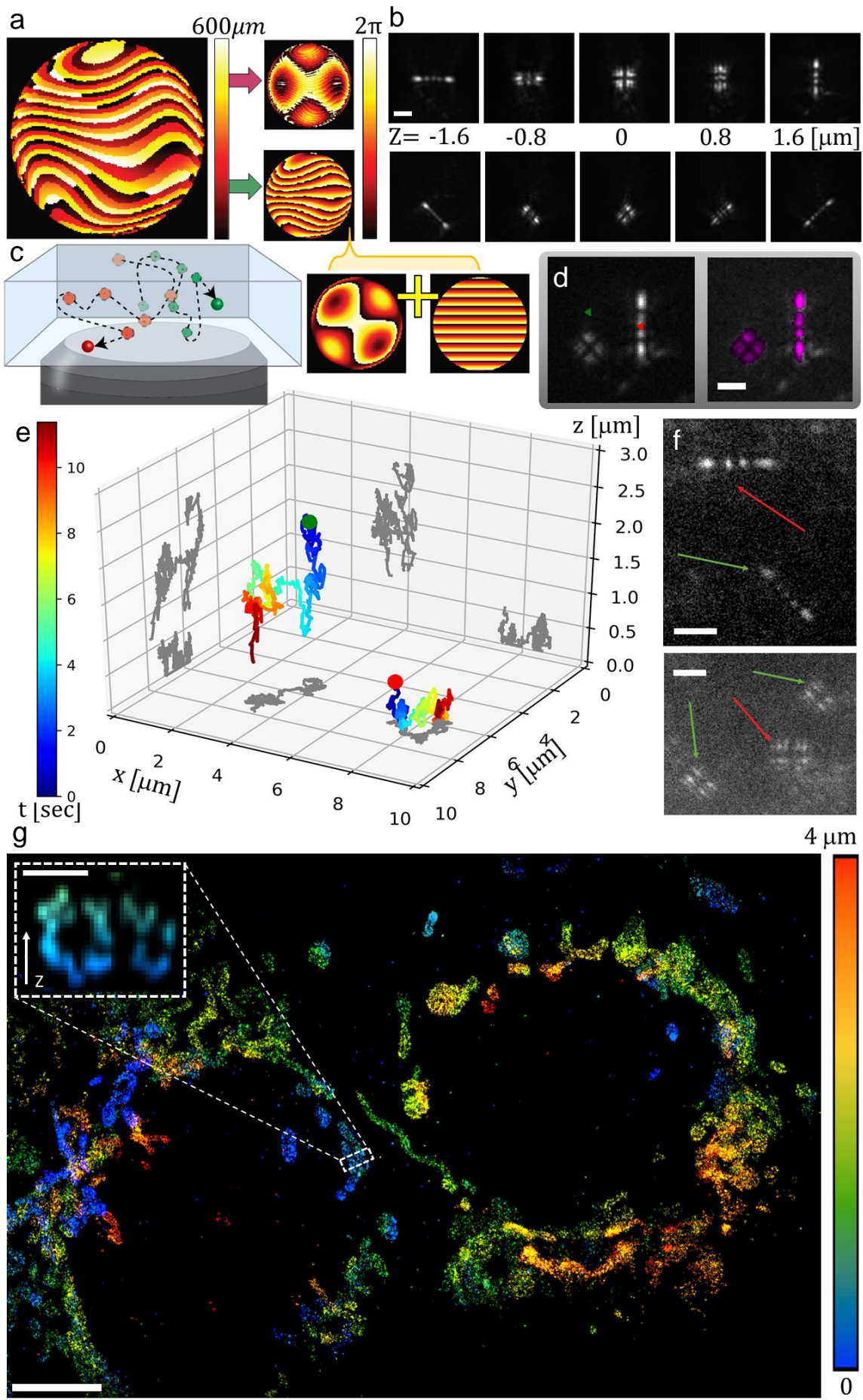

**Fig 6: Multicolor PSF-engineering DOE.** a) A heightmap of the multicolor DOE and corresponding phase patterns at 515nm and 680nm. b) Experimental images of fluorescent microspheres in two different colors at different axial positions using the same DOE c) Freely diffusing fluorescent microspheres in two different colors (emission peaks 515nm and 680nm) d) An example frame with the localizations obtained by the neural-net marked with red and green triangles (left) overlay with neural-net reconstruction in purple (right) e) The reconstructed 3D trajectories of the green and red fluorescent microspheres f) Single fluorophores in two colors (anti-mouse-AF488 and anti-mouse-AF647 antibodies) on a poly-lysine coated coverslip imaged simultaneously at two different axial positions g) Super-resolution STORM reconstruction of mitochondria in fixed COS7 cells labeled with anti-TOMM20-AF647 antibody, dashed zoomed-in cross-section: an example of a hollow mitochondria structure. Scalebar b,d and e: $2 \mu m$, g: $5 \mu m$, dashed cross-section: $0.5 \mu m$

To demonstrate the performance of the multicolor phase mask we performed two experiments. First, we tracked freely diffusing fluorescent microspheres in two different colors (peak emissions around 515 nm and 680 nm), color and depth-encoded by the mask, on a single optical channel. Figure 6f shows the 3D reconstructions of two 500-frame trajectories. For demonstrating the applicability of our multi-color mask under challenging single-molecule level SNR conditions, we performed 3D stochastic optical reconstruction microscopy (STORM)[36]. In this experiment, we labeled African green monkey kidney fibroblast-like (COS7) cells with the mitochondria-binding antibody TOMM20, directly conjugated to Alexa Fluor 647. The sample was illuminated with a 640 nm laser and a video containing frames of blinking fluorescent molecules was acquired. The reconstruction, namely 3D single-molecule localization was done by Deep-STORM3D[37] (Fig. 6g). Finally, to demonstrate the ability to image both red and green single fluorescent molecules simultaneously, we attached anti-mouse-AF488 and anti-mouse-AF647 antibodies to a poly-lysine coated coverslip and imaged them by simultaneously illuminating the sample with 488nm and 640nm lasers (Fig. 6e).

## Discussion

We demonstrate a simple and affordable method to fabricate high-quality solid DOEs using additive manufacturing combined with near index-matching. The process uses templates that can be fabricated at an easily obtainable resolution of tens of micrometers and achieves phase modulation equivalent to DOEs fabricated at a resolution which is orders of magnitudes higher. The process trades-off the nanoscale fabrication precision requirements with refractive index precision on the order of $4^{th}$ digit; the latter is typically easier to obtain.

In the context of computational imaging, where imaging systems are being co-designed alongside analysis algorithms, efforts are already being made on the software side to enable democratization of such methods[38]; the current work constitutes an important step towards increasing the availability of computational imaging methods from the hardware-side, specifically DOE, which has been typically a bottleneck. Fabrication cost of a DOE using this method is in the range of tens of dollars, and currently takes ~2-3 days, which mostly constitute waiting for room-temperature solidification; this step could be dramatically accelerated and will be addressed in future work, along with investigation of materials exhibiting different optical properties to serve as substrates for the DOE.

## Methods

### DOE fabrication

**1. Fabricating a microscale mold:** A ceramic mold that contains the inverse pattern of the desired DOE and a supporting base, was printed via additive manufacturing (Xjet Ltd., Carmel 1400, or Lithoz, CeraFab printer with ~25 µm resolution).

**2. Transferring the mold pattern for obtaining the first transparent layer:** The ceramic mold was first coated with a thin layer of oil (WD40-silicone) to reduce adhesion. Next, the first polymer was prepared and degassed in a standard vacuum chamber (Tarson, 402020) for ~10 minutes to remove bubbles. The first polymer was then poured onto the lubricated mold and placed into the vacuum chamber to facilitate additional bubble removal. The first layer was allowed to cure for 24 hours at room temperature on a leveled surface to ensure flatness. The polymerized first layer was then carefully separated from the ceramic mold, and the edges removed to achieve the desired size of the first layer. Then a brief oxygen plasma treatment (duration ~1 min, power ~7, Diener Electronic, Zepto W6) was done to the flat side of the first layer and to a high optical quality glass (Siegert Wafer, fused silica substrate, 35X35 mm$^2$ with a thickness of $500 \mu m$), then the two components were attached and placed in an oven at 70℃ for 30 minutes for bonding.

**3. Casting the second near index matched polymer:** We placed the first layer on a flat surface with the desired DOE profile facing up. Next, we prepared the second polymer (with near index matched refractive index) and degassed the solution in a vacuum chamber. Next, we filled up the entire dent (surrounding the desired profile) that was created from the supporting base and again placed it in the vacuum chamber to facilitate additional bubble removal for few minutes. Finally, the liquid polymer was carefully closed with second high optical quality glass from above and allowed to fully polymerized for 24-48 hours at room temperature.

**4. Positioning the DOE in the optical system:** Our element requires the same alignment procedure as standard DOEs for placing it in the desired plane of the optical system. For fine alignment, we attached a threaded adapter to the DOE (SM1S10, Thorlabs) with optical adhesive (NOA68T, Norland Products) and mounted it on a 6-axis kinematic optic mount (Thorlabs K6XS).

**Refractometry:** Polymer refractive indices were measured at seven different wavelengths (435.8nm, 480nm, 532nm, 589.3nm, 632.8nm, 700.1nm, 780nm) using a multiwavelength refractometer (Anton Paar, Abbemat MW) for achieving a high precision estimation for $\Delta n$ (up to fourth digit).

## STORM and single molecule experiments

**Cover glass cleaning for single molecule and STORM experiments:** 22X22 mm, 170 µm cover glasses (Deckgläser, No.1.5H) were cleaned in an ultrasonic bath with 5% Contrad 70 (Decon) at 60°C for 30 min, then washed twice with double distilled water (incubated shaking for 10 min each time), incubated shaking in ethanol absolute for 30 min, sterilized with filtered 70% ethanol for 30 min and dried in a biological cabinet.

**Single molecule sample preparation**: A clean cover glass was coated with poly-l-lysine by placing it in a 50 ml falcon containing 15 ml 0.01% poly-l-lysine solution (Sigma, P8920, diluted) for 10 minutes, then washing it twice with double distilled water and drying it in a chemical cabinet. An antibody solution was prepared by diluting anti-mouse-AF647 (Abcam, ab150115) and anti-mouse-AF488 (Abcam, ab150113) antibodies, 1:1000 and 1:100,000, respectively, in PBS. To attach the antibodies to the poly-l-lysine coated cover glass, we used a spin coater (Laurell) and apply 10 µl of the antibody solution onto the cover glass while spinning at 2,500 rpm for 35 sec. The cover glass was washed with 2 ml PBS and dried while spinning at 7,000 rpm for 1 min. We then applied a Gene Frame sticker (Thermo Fisher, AB0576), filled it with 25 µl PBS and stack a cover glass on top.

**STORM sample preparation**: COS7 cells at a concentration of 20,000 cells/ml in Dulbecco's Modified Eagle Medium (DMEM) with 1g/l D-glucose (Sartorius, 01-050-1A), supplemented with fetal bovine serum (Biological Industries, 04-007-1A), penicillin-streptomycin (Biological Industries, 03-031-1B) and glutamine (Biological Industries, 03-020-1B), were grown for 24 hr in a 6-well plate (Thermo Fisher, Nunclon Delta Surface) containing 6 ml of the cell suspension and the cleaned cover glasses, at 37°C, and 5% $CO_2$. The cells were fixed with 4% paraformaldehyde and 0.2% glutaraldehyde in PBS, pH 6.2, for 60 min, washed and incubated in 0.3M glycine/PBS solution for 10 minutes. The cover glasses were transferred into a clean 6-well plate and incubated in a blocking solution for 2 hr (10% goat serum, 3% BSA, 2.2% glycine, and 0.1% Triton-X in PBS, filtered with 0.45 um Millex PVDF filter unit). The cells were then immune-stained with 1:500 diluted anti-TOMM20-AF647 antibody (Abcam, ab209606) in the blocking buffer for 1.5 hr and washed five times with PBS.

For super-resolution imaging, a PDMS chamber (22x22x3 mm, with a 13x13 mm hole cut in the middle) was attached to the cover glass containing the fixed and stained COS7 cells to create a pool for the blinking buffer. Blinking buffer (50 mM Cysteamine hydrochloride (Sigma, M6500), 20% sodium lactate solution (Sigma, L1375), and 3% OxyFluor (Sigma, SAE0059) in PBS, pH 8-8.5)[39] was added and a cover glass was placed on top while ensuring minimal air bubbles.

**Imaging**: We used the Nikon eclipse Ti2 inverted microscope equipped with N-STORM unit (Nikon), silicone-oil objective (Nikon, RS HP Plan Apo 100x/1.35 Sil WD), and a multi-bandpass dichroic (Semrock, Di03-R405-488-532-635-t3). The microscope was extended with 4f system (f=200 mm) containing a multicolor phase mask in the Fourier plane and a sCMOS camera (Teledyne Photometrics, Kinetix) for image acquisition.

For single molecule experiments we illuminated the sample with 640 nm and 488 nm lasers simultaneously up to 80 mW each (Toptica, iChrome MLE). For the STORM experiment we illuminated the sample with high power 640 nm laser (~50 kW/cm$^2$ at the sample plane) and filtered the emission light with an additional 650 nm long pass filter (Thorlabs, DMLP650).

## **MINSTED experiment**

**Cell Sample Preparation**: Human osteosarchoma cells (U-2 OS cells) were grown on coverslips in McCoy's medium (16600082, Thermo Fisher Scientific, Waltham, MA USA) with 10% (v/v) fetal bovine serum (S0615, Bio&SELL, Feucht/Nürnberg, Germany), 1% (v/v) sodium pyruvate (S8636, Sigma-Aldrich) and penicillin-streptomycin (P0781, Sigma-Aldrich). For fixation, the cells were treated with 8% (w/v) paraformaldehyde in PBS at 37°C for 5 min. The samples were permeabilized using 0.5% (v/v) Triton X-100 in PBS for 5 min and blocked with 2% (w/v) bovine serum albumin (BSA) in PBS for 10 min. After incubation with primary antibody against Tom22 coupled with FITC (Clone: 1C9-2, 130-124-227, Batch: 5220902791, Miltenyi Biotec) diluted 1:50 in 2% (w/v) BSA in PBS for 1h, the samples were washed with PBS, and incubated with secondary antibody bearing DNA docking strands (Anti-Mouse IgG Docking site 1, Massive Photonics) diluted 1:100 in 2% (w/v) BSA in PBS for 1h. Subsequently to a washing with PBS, the cells were incubated with polyvinylpyrrolidone shelled silver nanoplates (SPPN980, nanoComposix, San Diego, CA USA) for 1h and washed with PBS.

**Single-molecule and DNA origami sample preparation**: Single molecules and DNA origami samples were prepared according to the protocol described in[31].

**Single-molecule E-PSF measurements:** Single-molecule samples were raster-scanned with different STED powers and fields of view adapted to the STED power. The obtained single-molecule signals were overlayed and the obtained PSF was fitted with a 2D Gaussian model to obtain the FWHM of the E-PSF at a given STED power.

**MINSTED imaging:** DNA origami samples were mounted with Cy3B (15 nM) coupled to the 3' end of the DNA oligonucleotide (sequence: CTAGATGTAT, Metabion) in 200 µl oxygen-deprived reducing-oxidizing buffer[40]. The buffer consisted of 100 µl reducing-oxidizing buffer (10% (w/v) glycose, 12.5% (v/v) glycerol, 0.1 mM TCEP, 1 mM ascorbic acid) and 100 µl PBS supplemented with 2 µl oxygen removal enzyme mix (25 units pyranose oxidase (P4234, Sigma-Aldrich), 80 µl catalase (C100, Sigma-Aldrich) with 170 µl PBS), 1 µl 200 mM methyl viologen dichloride hydrate (856177, Sigma-Aldrich) and 75 mM magnesium chloride. Cellular samples were mounted with 5 nM of Cy3B coupled to the 3' end of the DNA oligonucleotide (sequence: CTAGATGTAT, Metabion) in PBS with 75 mM magnesium chloride.

The obtained data was processed as described previously[31] and rendered based on estimated localization precision.

### resPAINT experiment

**Sample preparation for resPAINT imaging:** Large DOF imaging using DHPSF was performed with resPAINT as described previously[34]. Briefly, Jurkat T cells were fixed using 1% PFA for 30 min on ice, followed by 2x washes in PBS. Half of these cells were coated with fiducial markers to make 'bead cells'. This was done by mixing 50 µl of fixed cells with 1 µl of 100 nM 200 nm Deep Red FluoSpheres (F8810, Thermo Fisher) to let them spontaneously attach for 5 minutes, followed by 3x washes with PBS to remove free beads. The remaining Jurkat T cells were labelled with 1 nM Cellmask Deep Red (C10046, Thermo Fisher) for 5 minutes to allow distinguishing unlabelled cells from bead cells, followed by washing 2x with PBS.

To prepare the imaging sample, 3 µl of bead cells and 3 µl of uncoated cells were deposited onto a poly-D-lysine (PDL)-coated #1.5 coverslip. After 10 minutes of attachment, 3 µl of 20% bovine serum albumin was added to block the PDL and to minimise unspecific binding of our PAINT label. After 5 minutes of blocking, 7 µl of PH 9.6 sodium carbonate-bicarbonate buffer and 1 µl of 20 nM WGA-HMSiR was added, after which the sample was sealed and ready for imaging.

**ResPAINT Imaging of apical T-cell surface**: WGA-HMSiR was imaged on the apical surface of Jurkat T cells using a 1.35 NA silicone immersion objective lens with continuous 640 nm excitation (~ 6.8 kW/cm$^2$) in HiLo configuration. Autofocusing and DHPSF fitting is described in previous work[34] The DHPSF calibration was collected in steps of 60 nm, controlled by a piezo-objective stage, using ~100 nm nanoparticles extracted from Zebra Mildliner fluorescent markers attached to a #1.5 glass cover slip. 3D renders were created using VISP[41].

# Appendix

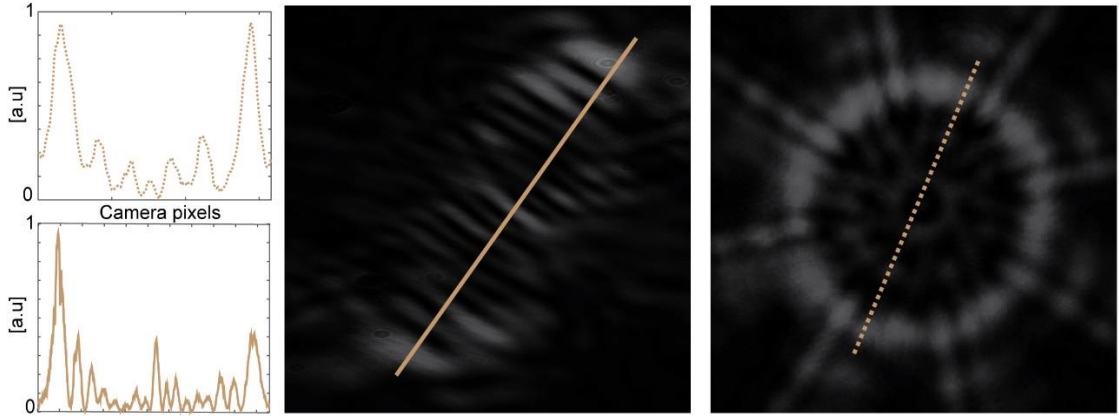

**Fig 7.** Experimental results of imaging a Gaussian beam through a SPP with m=16, using a cylindrical (left) and a spherical (right) lens.

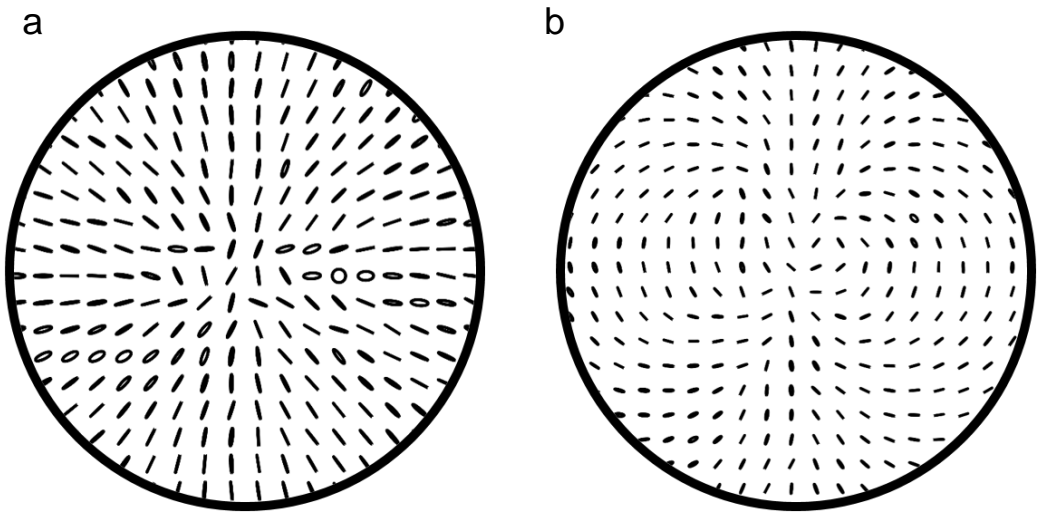

**Fig 8.** (a,b) distributions of the vector field reconstructed from the experimentally measured intensity profiles shown in Fig 3 (d,e).

## Acknowledgements

We would like to thank Michael S. Silverstein and Moran Bercovici for fruitful discussions, and Ophira Melamed, Yakir Tubul, Dov Grobgeld and Maor Yohanan (Xjet) and Peter Schneider and Johannes Homa (Lithoz) for help with optimization of the printing process. This work was funded by the European Union's Horizon 2020 research and innovation program under grant agreement No. 802567 -ERC- Five-Dimensional Localization Microscopy for Sub-Cellular Dynamics, under project number 101081911, HORIZON-ERC-POC, 3D-Optics, Israel Science Foundation, grant no. 969/22, a University of Leeds University Academic Fellowship, a Royal Society Research Grant (RGS\R2\202446) as well as an AMS Springboard Award (SBF006\1138) awarded to A.P.

# References

1.  O'Shea, D. C., Suleski, T. J., Kathman, A. D. & Prather, D. W. *Diffractive Optics: Design, Fabrication, and Test*. (SPIE, 2003). doi:10.1117/3.527861.

2.  Ries, H. & Leutz, R. Silicone-on-glass (SOG) diffractive optical elements (DOE) for the correction of chromatic aberrations and lens shape. in (2021). doi:10.1117/12.2595141.

3.  Drieschner, S., Kloiber, F., Hennemeyer, M., Klein, J. J. & Thesen, M. W. High quality diffractive optical elements (DOEs) using SMILE imprint technique. *Advanced Optical Technologies* **10**, (2021).

4.  Zhang, H., Liu, H., Xu, W. & Lu, Z. Large aperture diffractive optical telescope: A review. *Optics and Laser Technology* vol. 130 Preprint at https://doi.org/10.1016/j.optlastec.2020.106356 (2020).

5.  Antonov, A. I., Greisukh, G. I., Ezhov, E. G. & Stepanov, S. A. Diffractive elements for imaging optical systems. *Optoelectronics, Instrumentation and Data Processing* **53**, (2017).

6.  Yolalmaz, A. & Yuce, E. Designs of diffractive optical elements for solar energy harvesting. in *5th International Conference on Power Generation Systems and Renewable Energy Technologies, PGSRET 2019* (Institute of Electrical and Electronics Engineers Inc., 2019). doi:10.1109/PGSRET.2019.8882681.

7.  Shechtman, Y., Sahl, S. J., Backer, A. S. & Moerner, W. E. Optimal Point Spread Function Design for 3D Imaging. *Phys Rev Lett* **113**, 133902 (2014).

8.  Weber, M. *et al.* MINSTED fluorescence localization and nanoscopy. *Nature Photonics 2021 15:5* **15**, 361–366 (2021).

9.  Pavani, S. R. P. *et al.* Three-dimensional, single-molecule fluorescence imaging beyond the diffraction limit by using a double-helix point spread function. *Proceedings of the National Academy of Sciences* **106**, 2995–2999 (2009).

10. Tasso R. M. Sales & G. Michael Morris. Diffractive- and Micro-structured Optics. in *Review of Optical Manufacturing 2000 to 2020* 303–353 (2021).

11. Vijayakumar, A. & Bhattacharya, S. *Design and Fabrication of Diffractive Optical Elements with MATLAB* (2017). doi:10.1117/3.2261461.

12. Poleshchuk, A. G., Korolkov, V. P. & Nasyrov, R. K. Diffractive optical elements: fabrication and application. in *7th International Symposium on Advanced Optical Manufacturing and Testing Technologies: Design, Manufacturing, and Testing of Micro- and Nano-Optical Devices and Systems* vol. 9283 (2014).

13. Grushina, A. Direct-write grayscale lithography. *Advanced Optical Technologies* **8**, (2019).

14. Bhuvanesh Kumar, M. & Sathiya, P. Methods and materials for additive manufacturing: A critical review on advancements and challenges. *Thin-Walled Structures* vol. 159 Preprint at https://doi.org/10.1016/j.tws.2020.107228 (2021).

15. Camposeo, A., Persano, L., Farsari, M. & Pisignano, D. Additive Manufacturing: Applications and Directions in Photonics and Optoelectronics. *Advanced Optical Materials* vol. 7 Preprint at https://doi.org/10.1002/adom.201800419 (2019).

16. Zhang, D., Liu, X. & Qiu, J. 3D printing of glass by additive manufacturing techniques: a review. *Frontiers of Optoelectronics* vol. 14 Preprint at https://doi.org/10.1007/s12200-020-1009-z (2021).

17. Doualle, T., André, J.-C. & Gallais, L. 3D printing of silica glass through a multiphoton polymerization process. *Opt Lett* **46**, (2021).

18. Zhu, Y. *et al.* Recent advancements and applications in 3D printing of functional optics. *Additive Manufacturing* vol. 52 Preprint at https://doi.org/10.1016/j.addma.2022.102682 (2022).

19. Orange-Kedem, R. *et al.* 3D printable diffractive optical elements by liquid immersion. *Nat Commun* (2021) doi:10.1038/s41467-021-23279-6.

20. Shechtman, Y., Weiss, L. E., Backer, A. S., Lee, M. Y. & Moerner, W. E. Multicolour localization microscopy by point-spread-function engineering. *Nat Photonics* **10**, 590–594 (2016).

21. Maurer, C., Jesacher, A., Fürhapter, S., Bernet, S. & Ritsch-Marte, M. Tailoring of arbitrary optical vector beams. *New J Phys* **9**, (2007).

22. Zhang, Y. *et al.* Precision glass molding of diffractive optical elements with high surface quality. *Opt Lett* **45**, (2020).

23. Rahlves, M. *et al.* Flexible, fast, and low-cost production process for polymer based diffractive optics. *Opt Express* **23**, (2015).

24. Agarwal, A. K., Pant, K. K. & Mishra, S. K. Development of Microlens Arrays with Large Focal Number and High Fill Factor for Wavefront Sensing Applications. *IEEE Sens J* **21**, (2021).

25. Cai, S. *et al.* Microlenses arrays: Fabrication, materials, and applications. *Microscopy Research and Technique* vol. 84 Preprint at https://doi.org/10.1002/jemt.23818 (2021).

26. Zhang, Q. *et al.* Fabrication of Microlens Arrays with High Quality and High Fill Factor by Inkjet Printing. *Adv Opt Mater* **10**, (2022).

27. Volyar, A., Bretsko, M., Akimova, Ya. & Egorov, Yu. Measurement of the vortex and orbital angular momentum spectra with a single cylindrical lens. *Appl Opt* **58**, (2019).

28. Rosales-Guzmán, C., Ndagano, B. & Forbes, A. A review of complex vector light fields and their applications. *Journal of Optics (United Kingdom)* vol. 20 Preprint at https://doi.org/10.1088/2040-8986/aaeb7d (2018).

29. Hell, S. W. & Wichmann, J. Breaking the Diffraction Resolution Limit By Stimulated-Emission - Stimulated-Emission-Depletion Fluorescence Microscopy. *Opt Lett* **19**, 780–782 (1994).

30. Hell, S. W. Far-field optical nanoscopy. *Science* vol. 316 (2007) https://doi.org/10.1126/science.1137395.

31. Weber, M. *et al.* MINSTED nanoscopy enters the Ångström localization range. *Nat Biotechnol* (2022) doi:10.1038/s41587-022-01519-4.

32. Jungmann, R. *et al.* Single-molecule kinetics and super-resolution microscopy by fluorescence imaging of transient binding on DNA origami. *Nano Lett* **10**, (2010).

33. Lee, H. L. D., Sahl, S. J., Lew, M. D. & Moerner, W. E. The double-helix microscope super-resolves extended biological structures by localizing single blinking molecules in three dimensions with nanoscale precision. *Appl Phys Lett* **100**, (2012).

34. Sanders, E. W. *et al.* resPAINT: Accelerating Volumetric Super-Resolution Localisation Microscopy by Active Control of Probe Emission**. *Angewandte Chemie* **134**, 1–8 (2022).

35. Opatovski, N. *et al.* Multiplexed PSF Engineering for Three-Dimensional Multicolor Particle Tracking. *Nano Lett* **21**, 5888–5895 (2021).

36. Rust, M. J., Bates, M. & Zhuang, X. Sub-diffraction-limit imaging by stochastic optical reconstruction microscopy (STORM). *Nat Methods* **3**, 793–795 (2006).

37. Nehme, E. *et al.* DeepSTORM3D: dense 3D localization microscopy and PSF design by deep learning. *Nat Methods* **17**, 734–740 (2020).

38. von Chamier, L. *et al.* Democratising deep learning for microscopy with ZeroCostDL4Mic. *Nat Commun* **12**, 1–18 (2021).

39. Nahidiazar, L., Agronskaia, A. V., Broertjes, J., Van Broek, B. Den & Jalink, K. Optimizing imaging conditions for demanding multi-color super resolution localization microscopy. *PLoS One* **11**, (2016).

40. Vogelsang, J. *et al.* A reducing and oxidizing system minimizes photobleaching and blinking of fluorescent dyes. *Angewandte Chemie - International Edition* **47**, (2008).

41. El Beheiry, M. & Dahan, M. ViSP: Representing single-particle localizations in three dimensions. *Nature Methods* vol. 10 Preprint at https://doi.org/10.1038/nmeth.2566 (2013).

42. Hackel, T., Wegner, J. D. & Schindler, K. Contour detection in unstructured 3D point clouds. in *Proceedings of the IEEE Computer Society Conference on Computer Vision and Pattern Recognition* vols 2016-December (2016).